\documentclass[twocolumn,showpacs,prb]{revtex4}
\usepackage{bm}
\usepackage{graphicx}
\usepackage{amsmath}
\usepackage{bm}
\usepackage{epsfig}

\newcommand{\nix}[1]{}

\newcommand{\rmi}{{\rm i}}

\renewcommand{\Im}{\mathop\mathrm {Im}}
\renewcommand{\Re}{\mathop\mathrm{Re}}
\begin{document}

\title{Special frequencies in reflection spectra of Bragg multiple quantum well structures 
}
\author{M.~M.~Voronov, E.~L.~Ivchenko, A.~N.~Poddubny, V.~V.~Chaldyshev
} \affiliation{A.F.~Ioffe Physico-Technical Institute, Russian
Academy of Sciences, St. Petersburg 194021, Russia }
\begin{abstract}
We have studied theoretically optical reflection spectra from the
Bragg multiple quantum well structures. We give an analytical
explanation of the presence of two special frequencies in the
spectra at which the reflection coefficient weakly depends on the
quantum well number. The influence of the exciton nonradiative
damping on the reflection spectra has been analyzed. It has been
shown that allowance for the dielectric contrast gives rise to the
third special frequency at which the contributions to the
reflectivity related to the dielectric contrast and the exciton
resonance mutually compensate one another.
\end{abstract}
\maketitle

\section{Introduction}
The resonant Bragg structures have been first considered
theoretically in Ref.~\onlinecite{ftt} and then investigated
experimentally in systems based on semiconductor compounds
A$_2$B$_6$ \cite{kocher,merle,merle1} and A$_3$B$_5$
\cite{huebner,ell,hayes2,prineas}. The further progress in
understanding of optical properties of resonant Bragg structures
has been achieved in a series of theoretical studies
\cite{moiseeva,yakdr5,pss,deych1,kavokin,Ikawa_Cho,pilozzi,voronov}.
In resonant Bragg structures without the dielectric contrast,
i.e., with the coinciding dielectric constants, $\varepsilon_b
\equiv n_b^2$, of the barrier material and the background
dielectric constant, $\varepsilon_a \equiv n_a^2$, of the quantum
well (QW), the optical reflection spectrum for small enough number
$N$ of wells is described by a Lorentzian with the halfwidth $N
\Gamma_0 + \Gamma$, where $\Gamma_0$ and $\Gamma$ are,
respectively, exciton radiative and nonradiative damping
rates~\cite{ftt}. For a very large number of wells the reflection
coefficient is close to unity within the forbidden gap for exciton
polaritons and rapidly decreases near the gap edges $\omega_0 -
\Delta$ and $\omega_0 + \Delta$, where $\omega_0$ is the exciton
resonance frequency and $\Delta = \sqrt{2 \omega_0
\Gamma_0/\pi}$~\cite{pss,Ikawa_Cho}. In
Refs.~\cite{Ikawa_Cho,pilozzi} the reflection spectra are
calculated for arbitrary values of $N$, including the intermediate
region where $N \Gamma_0$ and $\Delta$ are comparable. The
calculations have demonstrated an existence of two particular
frequencies: near these special frequencies a value of the
reflection coefficient from the resonant Bragg structure with the
matched dielectric constants is practically independent of the QW
number and very close to that for reflection from a semi-infinite
homogeneous medium with the refractive index $n_b$. In the present
paper we give an analytical interpretation for this effect.
Moreover, we analyze the role of dielectric contrast
$\varepsilon_a - \varepsilon_b \neq 0$ in the formation of the
special frequencies in the reflection spectra.
\section{Structures with matched dielectric constants}
The structure under consideration is schematically depicted in
Fig.~1: it borders vacuum in the left and contains the cap layer
of the thickness $b'$ made from the material B, $N$ QWs made from
the material A, each of the width $a$, separated by the barriers
of the thickness $b$ and the semi-infinite medium B. Under normal
incidence (from vacuum) of the light wave of the frequency
$\omega$, the amplitude reflection coefficient can be written in
the following form~\cite{ftt}
\begin{equation}
\label{r} r(N) = \frac{r_{01}+\tilde{r}_N {\rm e}^{2 {\rm i}
\phi}} {1 + r_{01} \tilde{r}_N {\rm e}^{2 {\rm i} \phi}}\:.
\end{equation}
Here $r_{01} = (1 - n_b)/(1 + n_b)$ is the reflection coefficient
``vacuum $-$ semi-infinite medium B'', $\phi = k_b (b'-b/2)$, $k_b
= n_b(\omega/c)$, $c$ is the light velocity in vacuum,
$\tilde{r}_N$ is the reflection coefficient from the structure
with $N$ QWs placed between the infinite barriers. It is
convenient to refer the phase of the latter coefficient to the
plane shifted by $(a + b)/2$ from the center of the leftmost QW.
Then this coefficient is given by \cite{yakdr5,book}
\begin{equation}
\label{rNtilde} \tilde{r}_N=\frac{\tilde r_1}{1-\tilde
t_1\frac{\sin(N-1)Kd}{\sin NKd}}\:,
\end{equation}
where the complex coefficients $\tilde{r}_1, \tilde{t}_1$ describe
the reflection from and transmission through the layer of the
thickness $d = a + b$ containing a QW in its center, $K$ is the
wave vector of an exciton polariton at the frequency $\omega$ in
an infinite regular QW structure. For a structure without the
dielectric contrast one has \cite{book,pss}
\begin{gather}
\tilde{r}_1 = {\rm e}^{{\rm i} k_b d} r_1\:,\quad
\tilde{t}_1 = {\rm e}^{{\rm i} k_b d}(1 + r_1)\:,\nonumber\\ \label{oneQW}  \quad r_1 =
\frac{{\rm i} \Gamma_0}{\omega_0-\omega- {\rm i}(\Gamma_0 +
\Gamma)}\:.
\end{gather}

\begin{figure}[t]
  \centering
    \includegraphics[width=.42\textwidth]{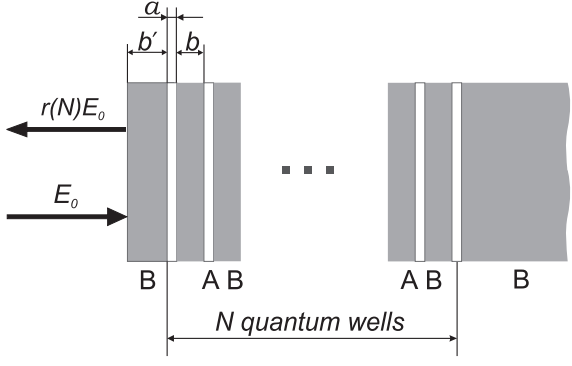}
  \caption{Schematic representation of light reflection from
   an $N$-QW structure. $E_0$ is the electric field amplitude of the
   incident light wave, and the reflection coefficient $r(N)$ is
   defined as a ratio of the reflected amplitude to $E_0$.
  }\label{f1}
\end{figure}

Fig.~2 shows the reflection spectra from the resonant Bragg
structure with matched dielectric constants. The spectra are
calculated in the absence of the nonradiative damping (a) and for
$\hbar \Gamma= 100\ {\rm \mu eV}$ (b). The rest parameters are
indicated in the figure caption. In agreement with
\cite{Ikawa_Cho,pilozzi}, one can see in the spectra presented in
Figs.~2a and 2b that for two frequencies labelled $\omega_+$ and
$\omega_-$ the reflection coefficient $R_N=|r(N)|^2$ is indeed
close to $r_{01}^2$ and almost independent of $N$, at least for
$N<100$. The special frequencies are tied to the exciton-polariton
forbidden-gap edges $\omega_0 \pm\Delta$ in such a way that
 \begin{gather}
\varepsilon_+\equiv  \omega_+-(\omega_0+\Delta)\ll \Delta\nonumber\\
\label{epsilon}\text {and }\:\: \varepsilon_-\equiv
(\omega_0-\Delta)-\omega_-\ll \Delta\:.
\end{gather}
On the other hand, the reflection coefficient $R_{\infty}$ from
the semi-infinite structure (curves $\infty$ in Figs.~2a and 2b)
shows abrupt change from values close to unity at $\Gamma \ne 0$
(or equal unity at $\Gamma \to + 0$) inside the forbidden gap to
values $R_{\infty}<r_{01}^2$ in the adjoining allowed bands. For
large but finite values of $N$ exceeding $1000$, the reflection
spectrum $R_N(\omega)$ from the structure with $\hbar \Gamma =
100\ {\rm \mu eV}$ is close to $R_{\infty}(\omega)$ while at
$\Gamma = 0$ the spectrum strongly oscillates outside the gap
region with the period decreasing as $N$ increases.
\begin{figure}[t]
  \centering
    \includegraphics[width=.48\textwidth]{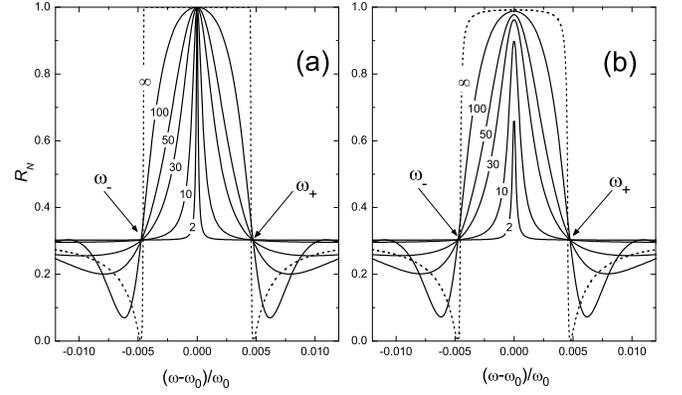}
  \caption{ Spectral dependence of the reflection coefficient
  $R_N$ from $N$-QW structure with the matched dielectric constants
  of compositional materials $A$ and $B$. The calculation is
  performed for the background refractive index $n_b = 3.45$,
  the exciton resonance frequency and radiative damping rate
defined by $\hbar \omega_0 = 1.533\ \rm{eV}$, $\hbar \Gamma_0=50\
\rm{\mu eV}$, $b' = (a/2) + b$ and the
nonradiative damping rate $\hbar \Gamma = 0$ (a) and
$\hbar\Gamma=100\ \rm{\mu eV}$ (b). Curves correspond to six
structures with different number of QWs indicated at each curve.
The reflection spectrum for the structure with infinite number of
QWs, $N \to \infty$, is labelled by the symbol $\infty$.
  }\label{f2}
\end{figure}

An existence of the special frequencies $\omega_{\pm}$ in the
reflection spectra can be understood if one notices that in the
vicinity of the edge frequencies $\omega_0 \pm \Delta$, where
\[
\left|N(Kd-\pi)\right| \ll 1,
\]
the ratio of the sine functions in \eqref{rNtilde} can be
approximated by
\begin{equation}\label{sinsin}
\frac{\sin(N-1)Kd}{\sin NKd}\approx-\frac{N-1}{N}\:.
\end{equation}
Therefore, the reflection coefficient $\tilde r_N$ and, hence, the
reflection coefficient \eqref{r} can be presented as the M\"obius
or linear-fractional transformation
\begin{equation}
\label{RFracLinear} r(N) = \frac{ \alpha N + \beta }{\gamma N +
\delta}
\end{equation}
with
\begin{eqnarray} \label{genabgd}
&& \alpha = r_{01} ( 1 + \tilde{t}_1 ) + {\rm e}^{2
{\rm i} \phi} \tilde{r}_1 \:,\: \beta = - r_{01} \tilde{t}_1\:,\\
&& \gamma = 1 + \tilde{t}_1 + {\rm e}^{2 {\rm i} \phi} r_{01}
\tilde{r}_1\:,\: \hspace{4 mm}\delta = - \tilde{t}_1\:. \nonumber
\end{eqnarray}
It should be noted that this equation is valid for an arbitrary
regular structure with $N$ periods if $\tilde{r}_1, \tilde{t}_1$
are considered as the reflection and transmission coefficients for
a single layer of the thickness $d$ embedded between the
semi-infinite media with the refractive index $n_b$. In the
particular case of an $N$-QW structure with the matched dielectric
constants we have instead of (\ref{genabgd})
\begin{eqnarray} \label{noabcd}
&&\alpha = r_{01} ( 1 + e^{{\rm i} \phi_d} ) (\omega_0 - \omega -
{\rm i} \Gamma) + {\rm i} \Gamma_0  (e^{{\rm i}(\phi_d + 2 \phi)}
- r_{01}) \:\nonumber,\:  \\ &&\beta = r_{01} \delta\:,\: \delta = -e^{{\rm
i} \phi_d} (\omega_0 - \omega - {\rm i} \Gamma)\:, 
\\ && \gamma = ( 1 + e^{{\rm i} \phi_d} ) (\omega_0 - \omega -
{\rm i} \Gamma) + {\rm i} \Gamma_0  ( r_{01} e^{{\rm i}(\phi_d + 2
\phi)} - 1) \:, \nonumber
\end{eqnarray}
where $\phi_d = k_b d$ and, for the sake of convenience, the
coefficients used here differ from those in (\ref{genabgd}) by the
common factor $\omega_0 - \omega - {\rm i} (\Gamma_0 + \Gamma)$,
which makes no change in the transformation
Eq.~(\ref{RFracLinear}).

Let us continue analytically the dependence $r(N)$ to the whole
complex plane ${z=z'+\rmi z''}$ and take into account that the
linear-fractional transformation $r(z)$ sends a circle to a
circle, a straight line can be considered as a circle of the
infinite radius, and the points ${z=1, 2, \ldots, N, \ldots}$ lie
on the real axis. It follows then that the complex values $r(N)$
lie on the circle of some radius $\rho$ centered at some point
$w_0$ so that one has
\begin{equation} \label{rnwr}
r(N) = w_0+\rho e^{\rmi\phi_N}\:,
\end{equation}
where only the phase $\phi_N$ is $N$-dependent. The values of
$w_0$ and $\rho$ are related to $\alpha, \beta, \gamma$ and
$\delta$ by \cite{lavshab}
\begin{equation}
\label{WandRho} w_0 = \frac {\rm i}{2}\ \frac{\alpha \delta^* -
\beta \gamma^*}{\Im(\gamma^* \delta)}\:,\quad \rho = \left|
\frac{\alpha}{\gamma} - w_0 \right|\:.
\end{equation}
According to (\ref{rnwr}) we have
\begin{equation}\label{Rn}
R_N \equiv |r(N)|^2 = |w_0|^2 + \rho^2 + 2 \rho \Re{(w_0^*
e^{\rmi\phi_N})}\:.
\end{equation}
If there exists a frequency $\omega$ where $w_0$ vanishes (or is
very close to zero) then the reflection coefficient $R_N$ at this
frequency is independent (or almost independent) of $N$ and equal
(or almost equal) to $|\alpha/\gamma|^2$.  If this frequency
satisfies the condition $|\omega -\omega_0|\gg \Gamma_0$, the
exciton contribution to the reflectivity is negligible for small
$N$ and values of $R_N$ are mainly determined by the reflection on
the boundary between vacuum and material B. Thus, for small $N$
the inequality ${|R_N(\omega)-r_{01}^2|\ll1}$ is valid. Since
$|w_0(\omega)|$ is negligible the same condition is satisfied not
only for small $N$ but for all those $N$ which allow the
representation of $r(N)$ in the form of linear-fractional
transformation \eqref{RFracLinear}.

The calculation shows that, for the structure characterized by the
parameters indicated in the caption to Fig.~2, the special
frequencies correspond to $\varepsilon_{\pm} \approx 0.045
\Delta$, and the ratio $|w_0(\omega) / r_{01}|$ reaches a minimal
value of $\approx 10^{-2}$ for $\Gamma = 0$ and $\approx 10^{-3}$
for $\hbar \Gamma = 100$ $\mu$eV. At the same time $\rho/r_{01}$
differs from unity less than by $10^{-3}$. In order to derive approximate analytical
expressions for the frequencies $\omega_{\pm}$ we can expand the
coefficients $\alpha, \beta, \gamma, \delta$ in \eqref{noabcd} in
powers of small parameters $\Delta/\omega_0, \Gamma_0/\Delta,
\varepsilon_-/\Delta$ or $\varepsilon_+ / \Delta$. Substituting
the obtained approximate equations into \eqref{WandRho} and
solving the equation $w_0=0$ we find
\begin{equation}\label{omegapmnocontr}
\omega_{\pm}=\omega_0\pm\Delta\left(1+\frac{1}{2n_b^2-3}\right)
\hspace{3 mm} \mbox{or} \hspace{3 mm} \varepsilon_{\pm} =
\frac{\Delta}{2 n_b^2 - 3}\:.
\end{equation}
For $n_b = 3.45$ the ratio $\varepsilon_{\pm}/\Delta$ following
from this equation differs from the numerical result only by
$8\%$. Therefore, the closeness of the frequencies $\omega_{\pm}$
to the gap edges is determined by a large value of the dielectric
constant $n_b^2$. We see that the approximation of $r(N)$ by a
linear-fractional transformation (\ref{RFracLinear}) and a
smallness of the minimal value of the function $|w_0(\omega)|$
allows one to explain the existence of special frequencies
$\omega_{\pm}$.

In contrast to the absolute value of the reflection coefficient
$r(N, \omega_{\pm})$ which is practically independent of the QW
number, the phase of the reflected wave $\phi_N(\omega_{\pm})$
appreciably changes with increasing $N$ and is described with a
good accuracy by the linear function
\begin{equation}\label{phias}
\phi_N (\omega_{\pm})\approx \pi\pm\frac{4 n_b \Gamma_0 N }{\Delta
(n_b^2 - 1)}\:.
\end{equation}
In Fig.~3 squares show the exact values of $R_N$ at the frequency
$\omega_+$ and the solid curve shows the approximate dependence
$R_N(\omega_+)$ obtained by the substitution of \eqref{phias} into
\eqref{Rn}. One can see that the approximate formula reproduces
the results of numerical calculation with high accuracy.
\begin{figure}[t]
  \centering
    \includegraphics[width=.48\textwidth]{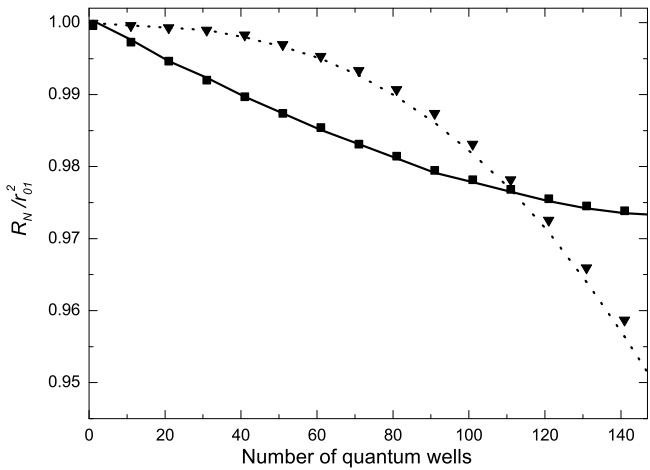}
  \caption{The dependence of the reflection coefficient
  $R(\omega)$ on the number of QWs at the frequency $\omega_+ =
\omega_0 + 1.045 \Delta$. The parameters coincide with those
indicated in the caption to Fig.~2. The exact values for the
structure with $\Gamma=0$ and $\hbar\Gamma = 100$ $\mu$eV are
shown by squares and triangles, respectively. Solid and dotted
curves represent the analytical approximation for $\Gamma=0$ and $
\hbar \Gamma = 100\ {\rm \mu eV}$.
  }\label{f3}
\end{figure}

Fig.~2b presents the spectra $R_N(\omega)$ calculated with
allowance for exciton nonradiative damping ${\hbar\Gamma=100\ {\rm
\mu eV}}$. One can see that the nonradiative damping leads to a
decrease of the reflection coefficient near the center of the
forbidden gap but makes no remarkable effect on the position of
the special frequencies $\omega_{\pm}$. Near these frequencies the
phase of the complex coefficient $r(N)$ is also described by
Eq.~\eqref{phias}. However the dependence of $R_N(\omega_{\pm})$
on $N$ for ${\hbar \Gamma =100\ \rm{\mu eV}}$ shown in Fig.~3 by
triangles essentially differs from that for $\Gamma=0$. For the
description of this dependence it is not enough to take into
account the variation of $\phi_N$ given by Eq.~\eqref{Rn}. The
reason is that, for nonzero nonradiative damping, the condition
for validity of the approximation \eqref{sinsin} is violated at
noticeably smaller $N$ than in the case of vanishing $\Gamma$. As
a result one has to keep the next nonvanishing term in the
expansion of the ratio of sine functions in powers of the variable
$Kd-\pi$, namely,
\begin{gather}
\frac{\sin{(N-1)Kd}}{\sin{NKd}} = - \frac{N-1}{N} F(N),\nonumber\\ \label{defF} F(N) =
1 + \frac{2N -1}{6}\ (Kd - \pi)^2\:.
\end{gather}
This expansion is valid for $N \le 100$. Since the factor $F(N)$
differs from unity the dependence $r(N)$ obtained in the
approximation (\ref{defF}) is not, strictly speaking,
linear-fractional: the coefficients $\alpha,\beta,\gamma,\delta$
become functions of $N$ and, therefore, the transformation $r(N)$
no more sends the real axis to a perfect circle. Formally, the
expression $r(N)$ can be presented in the form (\ref{rnwr}).
Moreover, the dependence of $\phi_N$ on $\Gamma$ can be neglected.
However, both $w_0$ and the real factor $\rho$ are now functions
of $N$. Triangles and dotted curve in Fig.~3 present results of
exact and approximate calculations of the coefficient $R_N$ at the
frequency $\omega_+$ for $\hbar \Gamma =$ 100 $\mu$eV. The
approximate calculation is carried out according to
Eq.~(\ref{Rn}), but with modified $w_0$ and $\rho$. For nonzero
nonradiative-damping rate, values of $K$ become complex even in
the allowed bands. The analysis shows that it is the imaginary
part of $K$ which gives rise to the difference between the solid
and dotted curves in Fig.~3.

\section{Allowance for the dielectric contrast}
For $n_a \neq n_b$, the reflection and transmission coefficients
for a single QW are described by \cite{yakdr5}:

\begin{gather}
\tilde r_1=e^{\rmi k_b d} r_1\:, \tilde t_1=e^{\rmi k_b d} t_1\:\nonumber,\\
 r_1 = r^{(0)} + r_{\rm exc}\:,\:t_1 = t^{(0)} + r_{\rm exc} \:.
\label{r1t1}
\end{gather}
Here $r^{(0)}$ and $t^{(0)}$ are the reflection and transmission
coefficients calculated neglecting the exciton contribution and
given as follows
\begin{gather}
r^{(0)} = e^{- {\rm i} k_b a} r_{ba} \frac{1 - e^{2 {\rm i}k_a
a}}{1 - r_{ba}^2 e^{2{\rm i} k_a a}}\:\nonumber,\\\: t^{(0)} = e^{{\rm i} k_a
a} \left(e^{- {\rm i} k_b a} + r_{ab}\: r^{(0)}\right)\:,
\label{r0t0}
\end{gather}
where $r_{ab} = - r_{ba} = (n_a - n_b)/(n_a + n_b)$. The exciton
contribution to $r_1$ and $t_1$ has the form
\begin{equation}
r_{\rm exc} = t^{(0)} \frac{{\rm i} \bar{\Gamma}_0}{\omega_0 -
\omega - {\rm i} \left( \Gamma + \bar{\Gamma}_0 \right)} \:,
\label{rexc}
\end{equation}
where
\begin{equation}
\label{gamma0} \bar{\Gamma}_0 = \frac{1 + r_{ab} e^{{\rm i} k_a
a}}{1 - r_{ab} e^{{\rm i} k_a a}}\: \Gamma_0 \:.
\end{equation}
As well as in the case $n_a = n_b$, the reflection coefficient
from an $N$-QW structure is expressed via those for a single QW
according to Eqs.~\eqref{r}, \eqref{rNtilde}. The period of the
structure is determined by the extended Bragg condition
\cite{yakdr5,voronov} which, for $|n_a-n_b| \ll n_b$, reduces with
high accuracy to
\[
\frac{\omega_0}{c}\ n_b d = \pi\:,
\]
which formally coincides with the exact condition for structures
with the matched dielectric constants.

\begin{figure}[t]
  \centering
    \includegraphics[width=.48\textwidth]{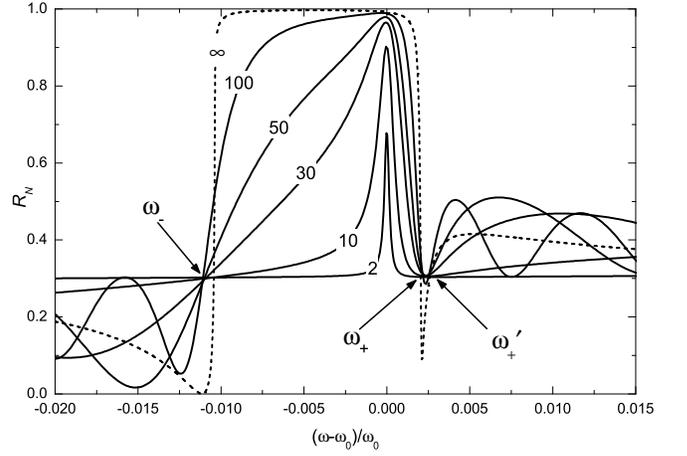}
  \caption{The reflection spectra from Bragg QW structures with
  the dielectric contrast between the compositional materials $A$
  and $B$. The calculation is performed for $\hbar \Gamma=100$ $\mu$eV,
  $a=120$ \AA, $n_a=3.59$ and $n_b = 3.45$. Curves are calculated
  for six structures containing different number, $N$, of wells indicated at
  each curve. The symbol $\infty$ corresponds to the structure
  with infinite $N$.
  }\label{f4}
\end{figure}

Fig.~4 shows the $N$-dependence of reflection spectra from the
resonant Bragg structures with the different refractive indices
$n_a$ and $n_b$. The values of parameters are indicated in the
figure caption. The calculation is performed with allowance for
the nonradiative damping $\hbar \Gamma=100\ {\rm \mu eV}$. One can
see that there are not two but three special frequencies in the
reflection spectra. Arrows point at these frequencies labelled as
$\omega_-$, $\omega_+$ and $\omega'_+$. An appearance of the
special frequencies $\omega_{\pm}$ can be explained in terms of
the linear-fractional transformation as well as for the structures
without the dielectric contrast. Since these frequencies are
located near the forbidden-gap edges one can apply the
approximation (\ref{sinsin}). For $\Gamma \ne 0$, similarly to
\eqref{defF}, it is necessary to take into account a term
quadratic in the difference $Kd-\pi$. As in the above case $n_a =
n_b$, an absolute value of the reflection coefficient $r(N)$ at
the frequencies $\omega_{\pm}$ weakly depends on $N$ whereas the
phase appreciably varies with $N$. The dependence $\phi_N$ at a
frequency $\omega$ lying close to the upper or lower edge of the
polariton gap can approximately be described by
\[
\phi_N(\omega) = \pi + \left( \frac{\Gamma_0}{\omega - \omega_0} -
2 \pi r_{ab}\ \frac{a}{d} \right)\ \frac{4n_b N}{n_b^2-1}
\]
valid for $|\omega - \omega_0| \gg \Gamma_0, \Gamma$. In the
particular case $r_{ab} = 0$ this equation transforms into
Eq.~(\ref{phias}).

Unlike the frequencies $\omega_{\pm}$, at the frequency
$\omega'_+$ both the absolute value and the phase of the
reflection coefficient $r(N)$ are practically independent of $N$,
i.e., not only $R_N = |r(N)|^2 \approx r_{01}^2$ but also $r(N)
\approx r_{01}$. This happens because the reflection of light at
the frequency $\omega'_+$ from a single QW sandwiched between the
semi-infinite barriers is almost absent, namely, a sum of two
terms $r^{(0)}(\omega'_+)$ and $r_{\rm exc}(\omega'_+)$ in
\eqref{r1t1} is close to zero. In other words, the contributions
to the reflectivity of a QW due to the presence of the dielectric
contrast and the exciton resonance cancel each other \cite{erementchouk}. According to
\eqref{rNtilde} the absence of reflection from one QW brings with
it the vanishing reflection coefficient from $N$ such wells, i.e.,
$\tilde r_N(\omega^{\prime}_+)=0$. As a result, the amplitude
reflection coefficient from the whole structure $r(N,
\omega^{\prime}_+)$ equals $r_{01}$ and is independent of $N$.

The frequency $\omega^{\prime}_+$ satisfies the inequalities
$|\omega'_+ - \omega_0 | \gg \Gamma_0, \Gamma$. This allows one to
neglect the damping rates $\bar\Gamma_0$ and $\Gamma$ in the
denominator of the expression \eqref{rexc} for $r_{\rm exc}$.
Furthermore, because of closeness of the refractive indices $n_a$
and $n_b$ the inequality $|r_{ab}| \ll 1$ holds. Neglecting
corrections of the order $r_{ab}^2$ in equations for $r^{(0)}$,
$t^{(0)}$ and the difference of $\bar \Gamma_0$ from $\Gamma_0$ we
obtain that the condition $r_1(\omega^{\prime}_+)=0$ is satisfied
at the frequency
\begin{equation}
\omega^{\prime}_+ = \omega_0 + \frac{\Gamma_0}{2r_{ab}\sin
k^{(0)}_a a}\:, \label{omega'}
\end{equation}
where $k^{(0)}_a=n_a\omega_0/c$. Obviously, the inequalities
${|\omega^{\prime}_+ - \omega_0 |}\gg \Gamma_0, \Gamma$ are realized
for a small enough coefficient $r_{ab}$. In the limit $n_a \to
n_b$ when $r_{ab}\to 0$, a value of $\omega^{\prime}_+$ tends to
infinity, i.e., this special frequency is absent for structures
with the matched dielectric constants. For parameters of the
structure used while calculating the spectra in Fig.~4, the
frequencies $\omega_+$ and $\omega'_+$ accidentally turn out to be
very close to each other.

\section{Conclusion}
For the resonant Bragg structures with the matched dielectric
constants of the well and barrier materials, we have given an
analytical explanation of an existence of the special frequencies
$\omega_{\pm}$ in the optical reflection spectra at which the
reflection coefficient $R_N$ is close to $r_{01}^2$ and almost
independent of the number of QWs in the structure. Near these
frequencies the amplitude reflection coefficient $r(N)$ can be
approximately written in the form of an $N$-dependent
linear-fractional function which sends points on the real axis to
points in the complex plane lying on a circle centered near the
coordinate origin. This means that the reflection coefficient
$R(\omega_{\pm})=|r(\omega_{\pm})|^2$ is indeed almost independent
of $N$ although the phase of the reflected wave ${\rm arg}\: r(N)$
is quite well approximated by a linear function of $N$. We have
shown that, for nonzero exciton nonradiative damping $\Gamma$, the
reflection spectra are also characterized by special frequencies
but their values at these frequencies become more sensitive to
$N$.

In optical reflection spectra from resonant Bragg structures with
the dielectric contrast there are not two but three special
frequencies. The origin of two of them can be interpreted in terms
of a linear-fractional transformation in the same way as in the
case $n_a = n_b$. The reflection coefficient from a single QW put
between the semi-infinite barriers vanishes at the third frequency
$\omega'_+$ because the contributions to the reflectivity
resulting from the dielectric contrast and the exciton resonance
mutually compensate one another. The consequence is that, at the
frequency $\omega'_+$, the amplitude reflection coefficient from a
structure containing an arbitrary number of such wells equals
$r_{01}$ as if the structure contained no QWs at
all.

\acknowledgments{The work is supported by the programme of Russian Academy of Sci.
and the Russian Foundation for Basic Research (grant 05-02-17778). 
A.N.P. acknowledges the financial support by the ``Dynasty'' foundation -- ICFPM.}

\end{document}